\magnification\magstep1

{\bf Quantum mechanical modeling of the CNOT (XOR) gate}

\bigskip

\centerline{Miroljub Dugi\' c}

\bigskip

\centerline{Faculty of Science, Dept. Phys., P.O.Box 60,
34 000 Kragujevac, Yugoslavia}

\centerline{E-mail: dugic@uis0.uis.kg.ac.yu}

\bigskip

{\bf Abstract:} We consider the CNOT quantum gate as a 
physical action, i.e. as unitary in time evolution of the
two-qubit system. This points to the modeling of the interaction
Hamiltonian of the two-qubit system which would correspond 
to the CNOT transformation; the analysis
naturally generalizes to the Toffoli gate. Despite nonuniqueness
of the model of the interaction Hamiltonian, the analysis 
distinguishes that the interaction Hamiltonian does not posses
any global (rotational) symmetry. This forces us to conclude
that the direct (non-mediated) interaction in the two-qubit
system does not suffice for implementing the CNOT gate. 
I.e., so as to be able succesfully to implement
the CNOT transformation, a mediator (i.e. an external physical
system interacting with both of the qubits) is required.

\bigskip

{\bf 1. Introduction}

\bigskip

Here we pose the question of quantum mechnical modeling of the
CNOT (XOR) gate.

The physical background is rather obvious: if one should like
to physicaly implement the CNOT transfrormation of the two-qubit
states, the CNOT action must be considered as a physical
dynamics of the two-qubit (2Q) system. That is, quantum mechanically,
the CNOT action represents a dynamical change of the states
of the 2Q system.

For the {\it isolated} 2Q system, the evolution in time
(dynamics of the system) is governed by the Schrodinger law,
i.e. with the unitary in time evolution operator, $\hat U(t)$.
Therefore, the quantum modeling of the CNOT gate is a task of
modeling the Hamiltonian of the 2Q system, {\it so as to one may
write}:
$$\hat U_{CNOT} = \hat U(t), \eqno(1)$$

\noindent
where $\hat U_{CNOT}$ is the unitary-operator representation of 
the logicaly defined the CNOT transformation. 

{\it Physically}, the task (1) refers to the {\it practical,
experimental realisation} of the mathematicaly defined the
CNOT transformation.

\bigskip

{\bf 2. Quantum mechanical form of the CNOT transformation}

\bigskip

Usually, the CNOT (XOR) gate (transformation) is defined [1]
by the unitary matrix :
$$U_{CNOT} = \pmatrix{1 & 0 & 0 & 0 \cr
                            0 & 1 & 0 & 0 \cr
                            0 & 0 & 0 & 1 \cr
                            0 & 0 & 1 & 0 \cr}, \eqno (2)$$

\noindent
but {\it bearing in mind} that this representation refers to the
"standard (computational basis" $\{\vert i\rangle_{1z} \otimes
\vert j\rangle_{2z}, i,j = 0, 1\}$ of the 2Q system consisting of
the mutually identical qubits; the states
$\vert i \rangle_{\alpha z}, \alpha = 1, 2$ are the eigenstates of 
the $z$-projection(s) of the spin(s), $\hat S_{\alpha z}$:
$$\hat S_{\alpha z} \vert 0\rangle_{\alpha z} = 
{\hbar \over 2} \vert 0\rangle_{\alpha z}$$
$$\eqno (3)$$
$$\hat S_{\alpha z} \vert 1\rangle_{\alpha z} = - 
{\hbar \over 2} \vert 1\rangle_{\alpha z}$$

However, this representation is not very informative.

We shall start from the logical (physical) definition of the
CNOT transformation, obtaining its the {\it operator form},
$\hat U_{CNOT}$. This will be the basis for solving the
task eq. (1).

{\it Physically (logicaly)}, the CNOT gate is defined [1]
as follows:

{\it Acting on the states from the "computational basis"
(cf. above), it does not change the state of the first
("controlled") qubit, but reverses the state of the
second ("target") qubit iff the state of the first qubit
is} $\vert 1\rangle_{2z}$.

Formally, it reads:
$$\hat U_{CNOT} \vert 0\rangle_{1z} \vert j\rangle_{2z} =
\vert 0\rangle_{1z} \vert j\rangle_{2z}, \quad j = 0, 1$$
$$\eqno(4)$$
$$\hat U_{CNOT} \vert 1\rangle_{1z} \vert j\rangle_{2z} =
\vert 1\rangle_{1z} \vert \neg j\rangle_{2z}, \quad j = 0, 1$$

\noindent
where "$\neg j$" means "not j": "not 0" = 1, and "not 1" = 0;
we omit the sign of the "direct product", $\otimes$.

With some care, but without particular difficulties, one obtains
{\it unique operator form} of $U_{CNOT}$:
$$\hat U_{CNOT} = \hat P_{1z} \otimes \hat I_2 +
\hat P_{2z} \otimes \hat \sigma_{2x}, \eqno (5)$$

\noindent
where we used the well known equality:
$$\hat \sigma_x \vert j\rangle_z = \vert \neg j\rangle_z.
\eqno (6)$$

\noindent
and $\hat P_{1z} = \vert 0\rangle_{1z}
{1z}_\langle 0\vert$, $\hat P_{2z} = \vert 1\rangle_{1z}
{1z}_\langle 1\vert$.

The expression (5) is the main result of this section.

\bigskip

{\bf 3. The task}

\bigskip

Now, the task (1) reduces to obtaining equality:
$$\hat U(t) = \hat P_{1z} \otimes \hat I_2 +
\hat P_{2z} \otimes \hat \sigma_{2x}, \eqno (7)$$

\noindent
where $\hat U(t)$ represents the unitary in time evolution
operator of the 2Q system.

I.e., the task is to {\it construct a model of the Hamiltonian of 
the 2Q system, which satisfies}:
$$\imath \hbar {d\hat U(t) \over dt} = \hat H(t) \hat U(t),
\eqno (8)$$

\noindent
{\it so as to fulfill eq. (7)}.

{\it For simplicity}, and in accordance with the quantum measurement 
and the decoherence theory [2-4], we shall consider the interaction 
Hamiltonian as the dominant term in the Hamiltonian of the system. 
Then we consider
$$\hat U(t) \sim \hat U_{int}(t), \eqno (9)$$

\noindent
where $\hat U_{int}(t)$ is "generated" by $\hat H_{int}$. [Notice
that this simplification becomes exact in the "interaction picture",
where:
$$\imath \hbar {d\hat U_{int} \over dt} = \hat H_{int I}
\hat U_{int}, \eqno (10a) $$

\noindent
and
$$\hat H_{int I} \equiv \hat U_{\circ}^{\dag} \hat H_{int}
\hat U_{\circ}. \eqno (10b)$$]

So, our task is to find a model of $\hat H_{int}$, which would
satisfy:
$$\imath \hbar {d\hat U_{int} \over dt} = \hat H_{int}
\hat U_{int}, \eqno (11)$$

\noindent
but so that one may write (cf. eq. (7)):
$$\hat U_{int}(t) = \hat P_{1z} \otimes \hat I_2 +
\hat P_{2z} \otimes \hat \sigma_{2x}. \eqno (12)$$

\bigskip

{\bf 4. Doing the task}

\bigskip

Certainly, from eq. (11) it follows:
$$\hat U_{int}(t) = \exp\{-\imath \int\limits_0^t
\hat H_{int}(t') dt'/\hbar\}. \eqno (13)$$

Now one meets the next {\it problem}: the l.h.s. of eq. (12)
exhibits the time dependecne, while the r.h.s. does not.

This problem can be resolved in few ways. Instead of being
exhaustive, here we shall consider the {\it simplest model of
the time independent} interaction, so as one may easily overcome
the time dependence of the l.h.s. of eq. (13).

We admit that {\it duration of the interaction} is $\tau$, i.e.,
that
$$\hat H_{int} = \hat V \quad {\rm for} \quad t \in [0, \tau], 
{\rm otherwise} \hat H_{int} = 0. \eqno (14)$$.

Certainly, then (13) reads:
$$\hat U(t) = \exp\{-\imath \tau \hat V/\hbar\}, \quad
\tau - fixed \eqno (15)$$

\noindent
assuming that the effect of $\hat U(t)$ on the initial state of the 2Q 
system, is not completed before $t \approx \tau$. (After this time 
interval, the 2Q system evolves freely.)

So, our task reduces to modeling $\hat V$, so as to one may write:
$$\exp\{-\imath \tau \hat V/\hbar\} = 
\hat P_{1z} \otimes \hat I_2 +
\hat P_{2z} \otimes \hat \sigma_{2x}. \eqno (16)$$

\bigskip

{\bf 4.1 A model of} $\hat V$

\medskip

Notice: the r.h.s. of (16) is diagonalizable (it is "separable"
[4]) in the noncorrelated basis $\{\vert i \rangle_{1z} 
\vert j\rangle_{2x}, i,j=0, 1 \}$. So, the same must apply to
the l.h.s. of eq. (16).

The {\bf simplest form} of $\hat V$ which could fit this
requirement is:
$$\hat V = \hat A_1 \otimes \hat B_2, \eqno (17)$$

\noindent
assuming that:
$$_{1z}\langle i \vert \hat A_1 \vert j \rangle_{1z}
= A_i \delta_{ij}$$
$$\eqno (18)$$
$$_{2x}\langle m \vert \hat B_2 \vert n \rangle_{2x}
= B_m \delta_{mn}$$

\noindent
and $\vert m \rangle_{2x}$ represent the eigenstates of $\hat
\sigma_{2x}$.

Clearly, eq. (18) is equivalent with
$$[\hat A_1, \hat \sigma_{1z}] = 0,$$
$$\eqno (19)$$
$$[\hat B_2, \hat \sigma_{2x}] = 0,$$

\noindent
i.e. with
$$\hat A_1 = \sum_i A_i \vert i\rangle_{1z}  \quad _{1z}\langle i\vert$$
$$\eqno (20)$$
$$\hat B_2 = \sum_m B_m \vert m\rangle_{2x}  \quad _{2x}\langle m\vert$$

Now, we should choose $A_i$s and $B_m$s, so as to satisfy eq. (16).

\bigskip

{\bf 4.2 A model of} $\hat A_1$ {\bf and} $\hat B_2$

\medskip

Bearing in mind eq. (20), the l.h.s. of eq. (16) - cf. [ ] -
reads: 
$$\exp \{-\imath \tau \hat V/\hbar\} = 
\hat P_{1z} \otimes  \exp\{-\imath \tau A_1 \hat B_2/\hbar\} +
\hat P_{2z} \otimes  \exp\{-\imath \tau A_2 
\hat B_2/\hbar\}. \eqno (21)$$

When compared to eq. (16), it leads to:
$$\exp \{ -\imath \tau A_1 \hat B_2/\hbar\} = \hat I_2
\eqno (22a)$$
$$\exp \{ -\imath \tau A_2 \hat B_2/\hbar\} = \hat \sigma_{2x}
\eqno (22b)$$

The choice $A_1 = 0$ is obvious.

On the other side, eq. (20) suggests that
the l.h.s. of (22b) - cf. [4] - can be
written as
$$\exp\{ -\imath \tau A_2 B_1/\hbar\} \hat \pi_{1x} +
\exp\{ -\imath \tau A_2 B_2/\hbar\} \hat \pi_{2x}, \eqno (23a)$$

\noindent
while the r.h.s. reads :
$$\hat \pi_{1x} - \hat \pi_{2x}. \eqno (23b)$$

Equating (23a) and (23b) it follows that
$$\exp\{ -\imath \tau A_2 B_1/\hbar\} = 1, \eqno (24)$$
$$\exp\{ -\imath \tau A_2 B_2/\hbar\} = -1, \eqno (25)$$

\noindent
which directly implies:
$$B_1 = {n h \over \tau A_2}, \quad
B_2 = {(2m + 1) h \over 2\tau A_2}. \eqno (26)$$

\bigskip

{\bf 4.3 A model of} $\hat V$

\medskip

So one obtains:
$$\hat V = \hat P_{2z} \otimes [(nh / \tau) \hat \pi_{1x} +
((2m + 1)h/2\tau) \hat \pi_{2x}], \eqno (27)$$

for mutually independent integers, $m, n$, and $\tau$ fixed.

Now one may wonder if, for fixed $\tau$, the interaction may
diverse from the exact duration $\tau$. But this does not make any 
particular problem. Let us suppose that the real interaction duration
equals $\tau' = \tau \pm \epsilon$, $\epsilon \ll \tau$. Then
eq. (15) reads:
$$\hat V' = \exp\{ -(\imath \tau' /\hbar \tau)
\hat P_{2z} \otimes [nh \hat \pi_{1x} +
((2m + 1)h/2) \hat \pi_{2x}]\} =$$
$$= \hat U(t) \cdot \hat u(t), \eqno (28)$$

\noindent
where
$$\hat u(t) = \exp\{\mp (\imath \epsilon /\hbar \tau)
\hat P_{2z} \otimes [nh \hat \pi_{1x} +
((2m + 1)h/2) \hat \pi_{2x}] \}, \eqno(29)$$

\noindent
and, obviously: $\hat u(t) = \hat I + O(\epsilon /\tau)$.
So, $\hat V$ and $\hat V'$ {\it satisfy the approximation
criterion} [1]: $\hat V - \hat V' \sim O(\epsilon /\tau)$.

\bigskip

{\bf 4.4 The Toffoli gate}

\medskip

In full analogy one may obtain the quantum-mechanical model
of the Toffoli gate. But this will be ommited here.

\bigskip

{\bf 5. The symmetry considerations}

\bigskip

Here we pose the question of the symmetry group of the
interaction eq. (27). A bit of care is required with this regard:
whilst the states $\{\vert i\rangle\}$ of both the qubits can
physically be virtually arbitrary (e.g., the "ground", $\vert g\rangle$,
and "excited", $\vert e\rangle$) states, all the considerations are
formally equivalent with a spin-1/2 system. It particularly means
that the actual Hilbert space(s) reduces to a 2-dimensional
space, and the corresponding algebra is the well known SU(2) algebra.
And this notion points out the symmetry groups that should be
considered.

As with the spin-1/2 system, the transformations to be considered
reduce to the next two unitary groups:

(i) the qubits' exchange (the permutation group), and

(ii) the global rotations of the two-spin-1/2 system.

\noindent
I.e., we assume that all the other transformations (from the
Galilei, or Poincare group) {\it are not defined}.

By the very definition (cf. Section 2), the CNOT transformation
{\it clearly distinguishes between the two qubits}: the "the
first qubit" is usually referred to as the "controlled qubit",
while the "the second qubit" is usually referred to as the
"target qubit". No exchange of the qubits is allowable.

So, it remains to consider the rotations.

As it is well known, the {\it global} rotations are {\it generated}
by the elements, $\hat S_n$ (a projection of spin along $\vec n$), 
of the SU(2) algebra:
$$\hat S_n = \hat S_{1n} + \hat S_{2n}. \eqno (30)$$

\noindent
That is, {\it the global rotation} about an axis $\vec n$ by the angle
$\theta$ reads:
$$\hat R_{\vec \theta} = \exp(-\imath \hat S_n \theta/\hbar).
\eqno (31)$$

But this operator can always be written in the (obviously
separable [4]) form:
$$\hat R_{\vec \theta} = \hat R_{\vec \theta}^{(1)} \otimes 
\hat R_{\vec \theta}^{(2)}, \eqno (32)$$

\noindent
where
$$\hat R_{\vec \theta}^{(i)} = \exp(-\imath \hat S_{in} \theta/
\hbar). \eqno (33)$$

So, for eq. (27), the global-rotations-symmetry-requirement implies
(as it can be easily seen):
$$[\hat S_{1n}, \hat \sigma_{1z}] = 0, \eqno (34a)$$
$$[\hat S_{2n}, \hat \sigma_{2x}] = 0. \eqno (34b)$$

However, and this is the point to be emphasized, {\it this cannot
be fulfilled}; at least not without changing the definition 
eq. (3) (cf. Section 6).

This notion follows from {\bf the isomorphism} between the 
Hilberty spaces of the two qubits. Particularly, the isomorphism
implies equivalence of eqs. (34a,b) with:
$$[\hat S_{in}, \hat \sigma_{iz}] = 0, \eqno (35a)$$
$$[\hat S_{in}, \hat \sigma_{ix}] = 0, \eqno (35b)$$

\noindent
for both $i = 1, 2$ - which certainly cannot be fulfilled for the 
{\it qu}bits. So we conclude that $\hat H_{int}$ {\it does not
have any global symmetry}!

\bigskip

{\bf 5.1 The isolated systems}

\medskip

Throughout this paper we examine (cf. Introduction) the
two-qubit system as an isolated quantum system.

To this end, for an isolated quantum ("microscopic") system
it is practicaly a matter of principle that its Hamiltonian 
has at least one group of the global symmetry. [E.g., for an 
EPR pair, there is the full (e.g., rotational) symmetry
of the "pair". In the collission processes it is both theoretically
and experimentally verified the perfect energy (momentum) 
conservation. The same applies to the radiative processes;
just remind the unsuccessful trial [5] in establishing
the oposite.]

However, in Section 4 we have considered the two-qubit
system as an isolated system, but we have obtained that
$\hat H_{int}$ does not have any global symmetry -
which, also, directly follows from eq. (5). This produces
{\bf a contradiction}.

\bigskip

{\bf 5.2 The contradiction}

\medskip

It is worth emphasizing the above distinguished contradiction.

Physically, it is practically a matter of principle to deal
with a global-symmetry group of the Hamiltonian of an isolated
quantum system.

But, as regards the CNOT transformation, such a group {\it does not
exist}.

\bigskip

{\bf 5.3 More general transformations}

\medskip

In order to find a "symmetry group" of $\hat H_{int}$, 
eq. (27), one could look for the more general transformations.
Certainly, this "search" reduces to looking for the hermite-conjugate
"generators" of the "symmetry" transformations which would 
commute with $\hat H_{int}$.

Then one may show that the "generators" of the nontrivial
transformations appear as the linear combinations of the
next operators:
$$\hat I_1 \otimes \hat \sigma_{2x}, \quad \hat \sigma_{1z} \otimes 
\hat I_2, \quad \hat \sigma_{1z} \otimes \hat \sigma_{2x}.
\eqno (36)$$

But the corresponding transformations have {\it no physical
interpretation in terms of the global transformations} of the
2Q system.

So, there remains the above conclusion: $\hat H_{int}$ does
not have any global symmetry.

\bigskip

{\bf 6. A proposal for removing the contradiction}

\medskip

The contradiction can be "easily" removed {\it by abandoning the
initial assumption} - that the 2Q system is isolated.

Without details, {\it the idea for overcoming the contradiction}
is as follows: To consider the 2Q system as an {\it open} system,
each qubit separately interacting with a {\it "mediator"}, i.e.
with an external system whose interactions with the qubits, effectively,
lead to the change of the states of 2Q system, as defined by $U_{CNOT}$.

Certainly, then $U_{CNOT}$ requires re-interpretation: it does not
refer to an isolated quantum system, but {\it it represents a
net effect of interaction of the qubits with a "mediator"},
which mediates the qubits' mutual "interaction". Finally, since
the 2Q system is an open system, its dynamics is neither unitary, 
nor unique [6], and the net-effect-$U_{CNOT}$ follows after
ignoring the states of the "mediator", $M$.

\noindent
{\bf A REMARK:} It is important to note that the paradox can be
also removed by {\bf redefining the definition}, eq. (3) in
either of the two ways: (i) by redefining the states of, e.g., the
first qubit: instead the eigenstates of $\hat \sigma_{1z}$, one could
consider the eigenstates of $\hat \sigma_{1x}$, which would lead to:
$$\hat U_{CNOT} = \hat P_{1x} \otimes \hat I_2 +
\hat P_{2x} \otimes \hat \sigma_{2x},
\eqno (37)$$

\noindent
with obvious symmetry (rotation about $x$-axis), and/or (ii) relativizing
the isomorphism between the Hilbert state spaces of the qubits:
e.g., by considering the mutually nonidentical qubits; then
eqs. (35a,b) need not to follow from eqs. (34a,b), {\it for the isomorphism
bears ambiguity}, thus reducing the problem onto the above
point ""(i)".

Still, with this regard appear further problems: Whether the
redifinitions can be successfully implemented for an {\it array}
of $n \gg 1$ the qubits, especially with regard to the necessity
of the different preparations of the states of the qubits, in practice.
Finally, both proposals bear ambiguities (concerning the definitions of
the qubits' states, and concerning the isomorphism), which open
the question, e.g., why would one deal with eq. (37), instead of with eq.
(5)?

So, we conclude that the above remark, i.e. necessity of
{\it mediating the interaction} betweent the qubits, proves
physically a favourable solution of the paradox, really
overcoming the above mentioned ambiguities.

\bigskip

{\bf 7. Conclusion}

\bigskip

The CNOT transformation of the two-qubit system, considered as 
an isolated quantum system, cannot be justified. For it impies
nonexistence of any global symmetry for the isolated two-qubit
system.

We propose to extend this system by a "mediator", $M$, so as to
the whole, $1+2+M$ evolves unitary in time, but so that when
ignoring the state(s) of $M$, as the net effect of the evolution
appears the CNOT transformation. This proposal will be elaborated 
elsewhere.

\bigskip

{\bf References :}

\item{[1]}
J. Preskill, Lecture Notes, website 
{\it www.theory.caltech.edu/~preskill/ph229}; D. Aharonov,
"Quantum Computing", LANL archive, quant-ph 

\item{[2]}
J. von Neumann, "Mathematical Foundations of Quantum Mechanics",
Princeton University Press, Princeton, 1955

\item{[3]}
W. H. Zurek, Phys. Rev. {\bf D26} (1982) 1862;
Phys. Today, October 1991, p. 26

\item{[4]}
M. Dugi\' c, Physica Scripta {\bf 56} (1997) 560

\item{[5]}
N. Bohr, H. A. Kramers and J. C. Slater, Phys. Mag.
{\bf 47} (1924) 785

\item{[6]}
O. Kubler and H. D. Zeh, Ann. Phys. (N.Y.) {\bf 76}
(1973) 405

\end